\documentstyle[12pt,epsf]{article}
\def\gsim{\mathrel{\rlap {\raise.5ex\hbox{$ > $}}
{\lower.5ex\hbox{$\sim$}}}}
\def\lsim{\mathrel{\rlap {\raise.5ex\hbox{$ < $}}
{\lower.5ex\hbox{$\sim$}}}}

\newcommand{\pr}{\paragraph{}}
\newcommand{\be}{\begin{equation}}
\newcommand{\ee}{\end{equation}}
\newcommand{\bea}{\begin{eqnarray}}

\newcommand{\eea}{\end{eqnarray}}

\def\gappeq{\mathrel{\rlap {\raise.5ex\hbox{$>$}}
{\lower.5ex\hbox{$\sim$}}}}
 
\def\lappeq{\mathrel{\rlap{\raise.5ex\hbox{$<$}}
{\lower.5ex\hbox{$\sim$}}}}
 
\begin{document}
 
\begin{titlepage}
\begin{flushright}

NTUA--75/99 \\
OUTP--98--90P \\
UOA-NPPS--1/99 \\

hep-th/9903045 \\
\end{flushright}

\begin{centering}
\vspace{.1in}
{\large {\bf On `Graceful Exit' from inflationary phase in two-dimensional 
Liouville String Cosmology }} \\ 
\vspace{.2in}

{\bf G.A. Diamandis}$^{a}$, {\bf B.C. Georgalas}$^{a}$,
 {\bf N.E. Mavromatos}$^{b,c}$ 
and {\bf E.~Papantonopoulos}$^{d}$ \\

\vspace{.5in}
 
{\bf Abstract} \\
\vspace{.1in}
\end{centering}
{\small Within the context of a super-critical (Liouville) string, 
we discuss (target-space) two-dimensional 
string cosmology. A numerical analysis indicates that
the identification of time with the Liouville 
mode results in an expanding universe with matter 
which exhibits an inflationary phase, and 
`graceful exit' from it, tending asymptotically 
to a flat-metric fixed point.
This fixed point is characterized by a 
dilaton configuration which, depending on the 
initial conditions, 
either decreases linearly with the 
cosmic time, or is a finite constant.
This implies that, in contrast to the 
critical string case, 
the string 
coupling remains bounded during the exit from the inflationary phase, 
and, thus, the pertinent 
dynamics 
can be reliably described in terms of a tree-level string effective 
action. 
The r\^ole of matter in inducing such phenomena 
is emphasized. It is also interesting 
to note that the asymptotic value of the 
vacuum energy, which in  the $\sigma$-model framework is
identified with the `running' central charge deficit,
depends crucially 
on the set of initial conditions. Thus, although preliminary, 
this toy model 
seems to share 
all the features expected to characterize a 
phenomenologically acceptable cosmological string model.}

\vspace{0.5in}
\begin{flushleft}
$^{a}$ University of Athens, Physics Department, 
Nuclear and Particle Physics Section, 
Panepistimoupolis, Ilisia GR 157 71, Athens, Greece. \\
$^{b}$ University of Oxford, 
Department of Physics, Theoretical Physics,
1 Keble Road,
Oxford OX1 3NP, U.K. \\
$^{c}$ CERN, Theory Division, CH-1211 Geneva 23, Switzerland. \\
$^{d}$ National Technical University of Athens, 
Physics Department, Zografou Campus, GR 157 80, Athens, Greece. \\

\end{flushleft}

\end{titlepage}

\section{Introduction} 

There is a vast number of inflationary models in the literature
~\cite{inflation} which try to give a
satisfactory answer to the traditional cosmological problems as the
age,size and flatness of the universe.Almost all of these theories
relay on a potential of a scalar field the so called ``inflaton".
The properties of this potential determine at a large extent the
success of the model.The scalar potential can arise in various
theories depending on the taste of the authors. An intriguing aspect of
all inflationary theories to date is the way inflation ends,
the so-called `graceful exit' problem.There is currently a vast
literature on the subject~\cite{inflation},dealing with this problem
from both the theoretical and the phenomenological points of view.
\pr

In string theory there were a lot of attempts to develop a consistent
theory of string cosmology,in which inflation plays a dominant r\^ole.
There were found solutions to the non-linear $\sigma$-model equations
for the
graviton,dilaton and antisymmetric tensor in a Friedman-Robertson-Walker
(FRW) background.Then the FRW spaces were interpreted as
conformal field theories capable of being embedded in consistent
string theories~\cite{aben,dgl}.
The advandange of this string cosmological model
is,that at tree level, does not depend on a scalar potential,so any
inflationary solution will depend on the dynamics of the scalar
field,and on its couplings to the other massless string modes.
Unfortunately, these tree level $\beta$-functions of the background
fields,
do not have any inflationary solution.To get inflation,
these equations must be modified by a non-trivial dilaton
potential which is generated by string loop effects.
\pr
Veneziano and his collaborators~\cite{veneziano} have suggested
that the dilaton of (critical) string theory 
plays the r\^ole of the inflaton field of conventional
inflationary theories~\cite{inflation}. In an implementation of this
idea,they suggested a pre-big-bang cosmology in which the kinetic
energy of a massless dilaton drives the inflation toward a
singularity.Before the singularity is reached,stringy and/or
nonperturbative effects bring an end to the inflationary phase and a
transition occurs to the standard Friedman-Robertson-Walker (FRW) 
Universe.
Considerable effort has been focused on the details of the graceful
exit from the inflationary era.It is still not at all clear that this
can be done.The reason is that such dilaton scenaria require knowledge of the
precise underlying dynamics, in particular the dilaton potential $V(\Phi )$. 
The dilaton field is strictly massless in the critical-dimension 
string theories and one would naively expect 
that there is no dilaton potential. 
\pr
However, compactification to four dimensions,
as appropriate for any attempt to discuss realistic physics of the 
inflationary 
Universe, or 
inclusion of higher world-sheet topologies (string loops),
may lead to all sorts of complications, and non-trivial 
interactions for the dilaton field, including the possibility of 
spontaneous breaking of scale symmetry, leading to a (small) mass for $\Phi$. 
The inclusion of higher string topologies seems unavoidable in 
the framework of~\cite{veneziano} due to the fact that the 
space-time 
configurations involved are characterized by a curvature growth 
and a dilaton $\Phi$ increasing linearly with the cosmic time. 
Since the string coupling $g_s \sim e^\Phi$, it is evident that 
for large times $t \rightarrow \infty$ one enters the regime of 
strong coupling, which makes the framework of the 
lowest-order (tree level) effective 
action inadequate. However, even under the inclusion of higher loops, or 
the construction of exact conformal field theories 
to take into account non-perturbatively such 
higher topologies~\cite{kounnas}, one 
could not avoid such indefinetely-growing string coupling
situations~\cite{ven2}. The system was not attracted 
by fixed points characterized by a constant, 
or decreasing with time, dilaton  configurations corresponding 
to bounded string coupling, thereby making the 
computations based on perturbative string theory 
unreliable. Thus,  
the 
exit problem in such string cosmologies persists. 
\pr
In addition to the lack of knowledge on the precise four-dimensional dilaton 
dynamics, the inflationary period is an {\it out of equilibrium} process. 
Indeed, in many recent attempts~\cite{outequil}
to study inflation, the concept of a {\it stochastic} process governing 
the inflationary phase, began to emerge, which requires means of
non-equilibrium field theories. In the context of critical-dimension string 
theory, where the 
various space-time field configurations obey world-sheet conformal-invariance 
conditions~\cite{strings}, equivalent to classical solutions of
the low-energy field equations, it is rather hard to incorporate 
the non equilibrium dynamics of the ``rolling-down-the-hill'' phase of 
inflation.
\pr

In this work we will present a two-dimensional stringy cosmological 
toy model in which inflation relays on the dynamic evolution of scalar
fields, and the transition from the inflationary phase to a
FRW Universe occurs in a natural way as a result of the
time evolution of the theory. To implement such a scenario, we modify
the $\beta$-functions of the background fields of a critical string theory
in such a way so as to include friction terms for all the fields
involved. A consistent way to do that is to consider a non-critical
(supercritical) 
string theory and idendify the time with the (time-like)
Liouville mode~\cite{emn,emninfl}. 
As emphasized in \cite{emninfl}, such scenaria 
lead naturally to stochastic-type inflation~\cite{outequil}.   
In the context of the two-dimensional cosmological string model, 
it will be shown that
the flow of (Liouville) time is such that the system 
is attracted by a fixed point characterized by a flat 
metric and a  
dilaton that, depending on the initial 
conditions,  either decreases linearly 
with the cosmic time, or tends to a constant value. This 
situation implies that the string coupling 
remains bounded during the transition from the inflation
to the post-inflation period, in contrast to the 
critical string situation~\cite{veneziano,ven2}. 
We note here that to get a non-critical string we have to include
matter, the r\^ole of which is played here by the Tachyon field,
which in two-dimensional target-space times is a 
massless matter field.
\pr
The structure of the article is as follows:
We first review in brief the generic formalism of time as a Liouville 
mode~\cite{emn},
and we discuss the generalized conditions of Liouville - restored conformal
invariance. These replace the critical-string conformal (Weyl) invariance 
conditions resulting in the vanishing of the world-sheet 
generalized $\beta$-functions (Weyl anomaly coefficients) 
for the set of
$\sigma$-model background fields. Next we examine the specific case of 
two-dimensional string cosmology. We solve the generalized set of 
Liouville-restored conformal invariance conditions for backgrounds 
corresponding to the matter and spacetime metric fields. The cosmological 
nature of the problem requires that such backgrounds exhibit solely 
Liouville ($=$ time) dependence. We seek, and find, solutions for the metric 
field, which exhibit inflationary behaviour at a certain stage of Liouville 
evolution, and we pay particular attention to demonstrate a 
``graceful exit'' from the inflationary phase. Asymptotically in 
time, the spacetime becomes an ordinary flat universe.
There is also a linear expansion at a late stage of the Liouville evolution. 
As a byproduct of our analysis we note that the above cosmological model is 
characterized by a non-zero Liouville($=$ time)-dependent vacuum energy, 
which relaxes to an asymptotic time independent value. The asymptotic value 
of the cosmological constant depends on the initial conditions; there are 
cases where this asymptotic value vanishes.
\pr
Our solutions are at present numerical. However we consider it a non-trivial 
fact that consistent inflationary scenaria, with the right properties to 
account for a ``graceful exit'' from the exponential expansion phase, can 
be found, at least numerically,
within the framework of non-critical Liouville strings. It is 
highly non-trivial that the Liouville-restored conformal invariance 
conditions yield this behaviour upon the identification of the Liouville 
field with time~\cite{emn,kogan}. 
The existence of such solutions in the two-dimensional toy 
cosmological model we consider here 
is encouraging for the extension of such analyses to  
higher-target-space-dimensional string theories. This is left for future 
investigations. 
\pr
We also stress the fact that the Liouville-restored conformal 
invariance conditions differ (to lowest order $O(\alpha)'$ in the 
Regge-slope
expansion) from Einstein's equations for the metric, encountered in 
analogous treatments of critical string models. 
From a ``field-theoretic'' 
point of view, 
this is to be expected from 
the fact that non critical Liouville dynamics is
describing out-of-equilibrium (non-classical) physical 
processes~\cite{emn}, 
believed to characterize the inflationary Universe. 
This is an important generic feature of Liouville dynamics, in the picture 
where one identifies the Liouville mode with the target time~\cite{emn,kogan}.

\section{Liouville String formalism}

After this introduction, we  now proceed with the analysis of our pilot
two-dimensional cosmological model. We commence 
with a brief description of the Liouville-dressing procedure for
non-critical string, with the Liouville mode viewed as a local world-sheet 
renormalization group scale~\cite{emn}.
Consider a conformal $\sigma$-model, described by an action $S^*$ on the 
world-sheet $\Sigma$, which is deformed by (non conformal) deformations 
$\int_{\Sigma}g^iV_id^2\sigma$, with $V_i$ appropriate vertex operators.

\be     
S_g = S^* + \int_{\Sigma}g^iV_id^2\sigma
\label{sigma}
\ee

The non-conformal nature of the couplings $g^i$ implies that their 
(flat)world-sheet renormalization-group $\beta$-functions, $\beta^i$, are non 
vanishing. The generic structure of such $\beta$-functions, close to a fixed 
point, $\{g^i = 0\}$ reads:

\be
\beta^i = (h_i - 2)g^i + c^i_{jk}g^jg^k + o(g^3).
\label{fixed}
\ee

In the context of Liouville strings, world-sheet gravitational dressing is 
required. The ``gravitationally''-dressed couplings, $\lambda^i(g,t)$, 
which from our point of 
view correspond to renormalized couplings in a curved space, 
read to $O(g^2)$~\cite{ddk,liouville}:

\be
\lambda^i(g,t) = g^ie^{\alpha_it} + \frac{\pi}{Q \pm 
2\alpha_i}c^i_{jk}g^j g^kte^{\alpha_it} + O(g^3), \qquad
Q^2 = c-25 \nonumber
\label{renorm}
\ee
where $t$ is the (zero mode) of the Liouville mode, $Q^2$ is the central 
charge deficit, and $\alpha_i$ are the gravitational anomalous dimensions:

\be
\alpha_i(\alpha_i + Q) = h_i - 2 \qquad  
{\rm for} \qquad  c \ge 25 \nonumber
\label{anom}
\ee
Below we shall concentrate exclusively to the supercritical string case, 
$Q^2 \ge 0$, which from the point of view of identifying the Liouville mode 
with target time, corresponds to a Minkowskian signature spacetime 
manifold~\cite{aben}. 
\pr
Due to the renormalization (\ref{renorm}), the critical-string conformal 
invariance conditions, amounting to the vanishing of flat-space 
$\beta$-functions, are now substituted by:

\be
{\ddot \lambda}^i + Q{\dot \lambda}^i = -\beta^i(\lambda)  \qquad 
{\rm for}~ c \ge 25. \nonumber
\label{neweq}
\ee
where the notation $\beta^i(\lambda)$ denotes 
flat-world-sheet $\beta$-functions but with the formal substitution
$g^i \rightarrow \lambda^i(g,t)$.  
Note the minus sign in front of the flat-world-sheet $\beta$-functions 
$\beta^i$ in (\ref{neweq}), 
which is characteristic of the supercriticality of 
the string~\cite{ddk,liouville}. 
As we see later, for our two-dimensional string cosmology, the sign will be 
crucial for the existence of acceptable inflationary solutions 
demonstrating ``graceful exit'' from the exponential expansion phase.
Upon the identification of the Liouville mode with the target time
the dot denotes temporal derivative. 
\pr
An important comment we would like to make concerns the possibility
of deriving the set of equations (\ref{neweq}) from a target space
action. This issue has been discussed in the affirmative in ref.\cite{
emninfl}, where it was shown that
the set of equations (\ref{neweq}) 
satisfies the Helmholtz conditions for the
existence of an action in the
`space of couplings' $\{ g^i \}$ of the non-critical string.
Upon the identification of target time with the
Liouville mode~\cite{emn} this action becomes identical
with the target space action describing the off-shell
dynamics of the Liouville string. We should stress
the fact that the action is off shell, in the sense that
the on-shell conditions correspond to the vanishing
of the $\beta$-functions $\beta^i$.
In our case $\beta^i \ne 0$, and the identification 
of the Liouville mode with the target time  
implies  
that the space-time 
graviton $\beta$-function on the right-hand-side of (\ref{neweq}),
as well as other target-space tensorial structures, appearing inside the 
$\beta ^i$ functions for the various modes,   
contain temporal (Liouville) components as well. 
In this respect, our non-equilibrium Liouville string approach 
to the temporal evolution~\cite{aben} should be contrasted
with the naive interpetation of a Liouville string as 
a critical equilibrium string living in a space time 
with one extra  dimension. In that case 
the corresponding $\beta$ functions
of the Liouville-dressed theory  
would satisfy classical equations of motion. 
As mentioned above, 
in our approach the conditions describing  the 
restoration of conformal invariance by means of Liouville 
dressing are {\it not} to be intepreted as classical equations 
of motion of a string living in a space-time with one extra   
target dimension.  
Thus our analysis below should be 
distinguished from previous analyses on Liouville cosmology~\cite{schmid}.  
\pr

A generic feature of Liouville dynamics, is that in terms of the world-sheet 
action, the normalization of the Liouville kinetic term can always be 
arranged (by chosing appropriate counterterms) to correspond to a target 
spacetime of Friedman-Robertson-Walker (FRW) type; i.e. the time-like metric
component (under the assumption that the Liouville mode is time) is:

\be
G_{00} = -1.
\label{frwform}
\ee

We remind the reader that the Minkowskian signature is due to 
supercriticality ($Q^2 = c-25 \ge 0$) assumption. This will be understood 
in what follows.
\pr
Before proceeding to discuss our two-dimensional 
cosmology, we would like to make a final remark 
concerning the physical origin of 
deviations from criticality in string theories. 
This will hopefully shed more light in the 
physics underlying our scenario. Deviations from conformal invariance imply, 
from a target-space viewpoint, that the relevant background field is 
``off-shell'', i.e. is not a classical solution of some equations of motion. 
As argued in \cite{emn}, such a situation may be encountered in the context 
of effective low-energy theories of quantum gravity. Due to quantum 
fluctuations of the metric field, corresponding to microscopic event 
horizons (spacetime boundaries), there is the possibility for low-energy 
propagating matter to be trapped inside or on such boundaries. 
This can be 
explicitly demonstrated, for instance, in the case of a Dirichlet (D) brane 
representation of such defects on spacetime. A low energy closed string 
state can split into open strings with their ends attached to the D-brane. 
Such trapping processes cannot be measured by a low-energy observer who 
performs local scattering experiments ``asymptotically far'' from these 
microscopic horizons. From the observer's point of view, therefore, the 
appearance of singular metric fluctuations will result in making the 
low-energy matter system, consisting of propagating degrees of freedom only, 
an ``open'' quantum system. Information is ``lost'' into degrees of freedom 
pertaining to the quantum recoil of the D-brane horizon. In turn, such recoil 
degrees of freedom cause a sufficient distortion of spacetime, characterized 
by non-trivial particle creation, leading to decoherence of the low-energy 
matter system. A detailed analysis and formalism of such issues has been 
documented in \cite{kanti}, 
where we refer the interested reader for details. 
\pr
For our purposes below we 
assume that, within the context of a two-dimensional cosmological model, 
such quantum processes in the early Universe have resulted in a
matter deformation of the stringy $\sigma$-model, which does not satisfy 
classical equations of motion and hence, from a conformal point of view, 
departs from the critical string case. Back-reaction of such matter onto 
spacetime structure, prevents the metric from satisfying Einstein's 
equations (to lowest order in the $\alpha'$ expansion, where we concetrate 
ourselves throughout this work.)
\pr
With these in mind, our proposal for the two-dimensional string cosmology 
may now be 
formulated as follows: the low-energy (local, propagating) fields are the 
metric, 
$G_{ij}$, 
the 
dilaton, $\Phi $,  and ``tachyon'', $T$, fields~\footnote{Due to the Abelian 
Gauge symmetry 
$B_{MN} \rightarrow B_{MN} + \partial_{[M}\Lambda _{N]}$, in two-dimensional 
space times the antisymmetric tensor field
has no propagating modes, given that it may be eliminated from the 
low-energy action. The remaining discrete mode is thereby 
considered part of the unobservable
gravitational environment.}. In flat space times, the two-dimensional 
tachyon field is actualy massless and constitutes 
our low-energy matter. As we shall see the presence of such matter
is crucial for the inflationary scenario. 
\pr
The ${\cal O}(\alpha ')$ $\beta$-functions, 
corresponding to these fields, read:
\bea
{\rm Graviton} \qquad {\tilde \beta}_{ij}^G & =& \alpha ' \left(R_{ij} +
 2 \nabla _i \nabla _j \Phi - \nabla _i T \nabla _j T \right)\nonumber \\
{\rm Dilaton} \qquad {\tilde \beta}^\Phi & =& - R + 4 \nabla _i \Phi 
\nabla ^i \Phi - 4 \nabla ^2 \Phi + \nabla _i T \nabla ^i T - 2T^2 + Q^2 
\nonumber \\
{\rm Tachyon} \qquad {\tilde \beta}^T & =& - 2 \nabla ^2 T +
4\nabla _i \Phi \nabla ^i T - 4T.
\label{betaeqs}
\eea

In the above we have taken into account the freedom to
fix the tachyon potential
in string theory~\cite{banks}, by appropriate field redefinitions, such that
it only incorporates $T^2$ terms ($V(T)=-2T^2$).
The tilde denotes Weyl anomaly coefficients,
which replace the ordinary renormalization-group
$\beta$-functions in the case of stringy $\sigma$-models, as a result of 
target-space
local diffeomorphisms~\cite{schmid,tseytlinshore}. 
\pr
In the context of critical strings 
the vanishing of these ${\tilde \beta}$-functions can be interpreted as 
equations of motion from the action
\be
 S = \int d^2x \sqrt{-g} \left\{ e^{-2\Phi} \left[ R +
 4(\nabla \Phi)^2 - (\nabla T)^2 - V(T) - Q^2 \right] \right\},
\label{actioneff}
\ee
In this notation the string coupling is $g_s = e^{\Phi}$. 
As a general remark we stress that in two target-space-time dimensions
there is no Einstein frame, i.e. a frame in which the 
conformal dilaton factor $e^{-2\Phi}$ in front of the 
curvature term in the action (\ref{actioneff}) 
can be removed by a field redefinition. Thus, in this case 
the $\sigma$-model frame is also the `physical' frame. This 
will always be understood in the following.
\pr
In our non-critical string context the Weyl-anomaly coefficients
${\tilde \beta}$ are related off-shell with variations 
of the above action~\cite{osborn}, 
\be
{\tilde \beta}^i \sim {\cal G}^{ij}\frac{\delta S}{\delta g^j}
\label{offshell}
\ee
where ${\cal G}^{ij}$ is related to the (inverse) 
Zamolodchikov metric
in theory space~\cite{zam}, given by the world-sheet 
two-point function of vertex operators.
As mentioned previously, 
the set of equations (\ref{neweq},\ref{offshell})
satisfies the Helmholtz conditions for its
being derived from an off-shell action in theory space
of the non-critical string~\cite{emninfl}. 
The purpose of this article is to point out that, 
upon the identification 
of the Liouville mode with the target time, this {\it specific set} 
of field equations 
will 
yield cosmological solutions, capable of describing inflationary phase
of an expanding string universe and graceful exit from it.
\pr
To this end we combine (\ref{neweq}) with (\ref{betaeqs}), 
where now the indices $i,j=1,2$ span a two-dimensional target space time. 
The dots refer to (Liouville) 
time $t$ derivatives, and the cosmological model is obtained by assuming that
the various background fields exhibit only (Liouville) time $t$ dependence.
The metric is assumed to have the FRW form:
\be
G_{ij} = \left(\begin{array}{cc}
-1 \qquad 0 \nonumber \\0 \qquad e^{b(t)}\end{array}\right) 
\label{metricfrw}
\ee
\pr
The important comment we wish to make concerns the fact that, due to the 
renormalizability of the (non-critical) $\sigma$-model, there is an additional 
equation~\cite{curci} which should supplement (\ref{neweq}), the 
Curci-Paffutti 
relation, which relates the dilaton $\beta$ function, and hence the
effective running central charge of the theory, with the rest of 
the $\beta$ functions: 
\be
-\nabla _i {\tilde \beta}^\Phi + 2 G^{lj}\nabla _l\beta^G_{ij} - 
4 G^{lj}\nabla _l\Phi 
\beta^G_{ij} + \nabla _i {\tilde \beta}^T = 0
\label{pafuti}
\ee
Although this equation holds formally in the flat world-sheet case,
however in our framework it should also hold for the 
$\beta^i(\lambda)$ functions, i.e. the flat-world-sheet
$\beta$ functions upon the substitution of the 
$\sigma$-model couplings with the Liouville dressed ones. 
It will provide a highly non-trivial constraint, which should be respected
by the process of identifying the Liouville (world-sheet) 
renormalization scale 
with the target time~\cite{emn}. 
\pr
We are now seeking solutions to the system of equations (\ref{neweq}),
(\ref{betaeqs}),(\ref{pafuti}) exhibiting at a certain stage in Liouville 
time $t$ inflationary behaviour (exponential
expansion in the spatial volume). Given the choice for the metric
 (\ref{metricfrw}), and the assumption that all the fields are time
  dependent
 only, the equations (\ref{neweq}) take the form:
 \bea
 \frac{3}{2} \ddot{b} + \frac{5}{4} \dot{b}^2 + 
   \dot{b}(Q - \dot{\Phi}) &=& 0 \nonumber \\
 \frac{\ddot{b}}{2} + \frac{\dot{b}^2}{4} - 2 \ddot{\Phi} + 
 \dot{T}^2 &=& 0 \nonumber \\
 5 \ddot{\Phi} - \ddot{b} - \frac{1}{2} \dot{b}^2 - 
  4 \dot{\Phi}^2 + 
 \dot{\Phi}(Q + 2 \dot{b}) -
 \dot{T}^2 - 2 T^2 + Q^2 &=& 0 \nonumber \\
 3 \ddot{T} + \dot{T}(\dot{b} - 4 \dot{\Phi} + Q) -
   4T &=& 0 \nonumber \\
 \dot{\ddot{b}} - 4 \dot{\ddot{\Phi}} + \dot{b} \ddot{b} +
  8 \dot{\Phi}\ddot{\Phi} -  2\ddot{b} \dot{\Phi} -
 2\ddot{\Phi}\dot{b} + \dot{T}^2 (\dot{b} - 4 \dot{\Phi})+
 4 \ddot{T}\dot{T} - \dot{Q}Q &=& 0,
 \label{finaleq}
 \eea
 where the first four refer to the diagonal components of the
 metric, the dilaton and the tachyon fields and the last one
 is the time component of the Curci-Paffuti equation (\ref{pafuti}).
 Note that the equation for the non-diagonal component of the
 metric and the space component of the Curci-Paffuti equation
 are trivially satisfied.
\pr
It can be seen, for example, from the first of the equations
 in (\ref{finaleq})
that in principle there are both inflationary and non-inflationary solutions.
In particular if we assume that the dilaton and the central charge are slowly
varying with time, then  negative values of the quantity $( Q(t)-
\dot{\Phi}(t))$, gives an exponentially growing scale factor ($exp(b(t))$),
 while positive values of the same quantity yields power-law scale factor.
  Thus
 we have to see if our system of equations gives  the right relative
  magnitude
 to the dilaton and the central charge in order to ensure an inflationary era
 followed by a power-law expansion, leading asymptotically to flat space.
\pr
 Using the first the second and the fourth
 equations in (\ref{finaleq}) we solve for the second
 derivatives of the fields $b$, $\Phi$ and $T$, and eliminate their
 higher derivatives from the last one, concluding to the
 following set of four equations:
  \bea
  \ddot{b} &=& -\frac{2Q\dot{b}}{3} + \frac{2\dot{\Phi}~\dot{b}}{3} - 
  \frac{5\dot{b}^2}{6} \nonumber \\
  \ddot{\Phi}& =& -\frac{Q\dot{b}}{6} + \frac{\dot{\Phi}\dot{b}}{6} -
   \frac{\dot{b}^2}{12} + \frac{\dot{T}^2}{2} \nonumber \\
  \ddot{T} &=& \frac{4T}{3} - \frac{Q\dot{T}}{3} + 
  \frac{4\dot{\Phi}\dot{T}}{3} - \frac{\dot{b}\dot{T}}{3} \nonumber \\
  Q~\dot{Q} &=& \frac{Q\dot{b}^2}{3} + \frac{\dot{b}^3}{6} +
 \frac{2}{3}\dot{b}^2\dot{\Phi}.
   \label{system}
  \eea
 The equation coming from the dilaton $\beta$-function (third equation in
 (\ref{finaleq})) is left as a compatibility condition.

\section{Inflationary Solutions in Liouville String Theory}

 In order to get solutions 
 it seems necessary to distinguish two major cases: (i) solutions
where the central charge deficit $Q^2 (t) \rightarrow {\rm const} \ne 0$
as $t \rightarrow \infty$, and (ii) solutions where 
$Q^2 (t) \rightarrow 0$, as $t \rightarrow \infty$. 
Although, as we shall see, from an effective field theory 
view point, {\it both} cases seem plausible, however, at least
at present, 
it is not clear to us whether both situations can be met 
in an actual string theory framework. 
Indeed, in case (i), 
the asymptotic string theory will be a standard 
non-critical string theory in two-dimensional target-space, with the 
constant central charge deficit compensating the non-critical
dimensionality of the target space~\cite{ddk,aben}. 
Such string theories are known to exist as exact conformal field theory 
models. 
In case (ii), however, 
the resulting string theory will be a critical 
string theory. The asymptotic vanishing of the central charge,
which plays the r\^ole of a target-space cosmological constant, 
may be understood in that case as a result of the 
effects of the higher-level string modes of the target-space two-dimensional
string. At present, we do not have 
an exact conformal field theory description of such models, but in 
view of the existence of consistent non-trivial examples 
in the case of stringy black-holes~\cite{gang6}
we conjecture that such a case can also represent consistent  
cosmological string backgrounds.

\subsection{Solutions attracted by a linear-dilaton fixed point}

We start our analysis from case (i). To obtain the solutions
of (\ref{system}) in this case
we adopt the following "quasi-linear" method.
 We separate
  the fields in their asymptotic values plus fields which tend
  asymptotically to
 zero, namely:
 \bea
 \dot{\Phi} &\equiv d& + \phi_1 \nonumber \\
  Q &\equiv&  Q_0 + Q_1 \nonumber \\
 \dot{b} &\equiv & b_1 \nonumber \\
 T &\equiv T_0&, \qquad \dot{T} \equiv T_1,
 \label{separate}
 \eea 
 where the constants $d$ and $Q_0$ are related through the relation
 $Q_0 = -d(1+\sqrt{17})/2$, which results from the
 requirement  that the dilaton equation
 is satisfied,
 and the fields $(\phi_1,\:b_1,\:T_1,\:T_0,\:Q_1)$ vanish asymptotically.
 We assume that the constant $d < 0$ in order to have weak gravity
 asymptotically in time. (The case $d=0$ will be discussed separately).
  
The system can be written in the form
 \be
 \dot{\vec{x}} =  {\bf A} \vec{x} + {\bf \vec{F}}(\vec{x})
 \label{quasilinear}       
 \ee
 where $\vec{x} =  (\phi_1,\:b_1,\:T_1,\:T_0,\:Q_1)^{\bot}$, ${\bf A}$ is 
 the $5 \times 5$ matrix determining the linear part of the system and ${\bf 
 \vec{F}}(\vec{x})$ gives the nonlinear terms. 
 \newline
 The solution of the system can be given in an iterative form:
 \be
  \vec{x}_{(n+1)}(t) =  \vec{x}_{(n)}(t) + \int_{t_0}^{t}ds {\bf Y}(t)
 {\bf Y}^{-1}(s) \vec{{\bf F}} \left[ \vec{x}_{(n)}(s) \right],
 \label{itersol}
 \ee
 where the matrix ${\bf Y}$ satisfies the equation
 \be
 \dot{{\bf Y}} =  {\bf AY}. \nonumber
 \ee
  Note that the iteration in (\ref{itersol}) converges to the full
 solution.
 The starting point of the iteration procedure is the solution of
 the linear system with the correct asymptotic behaviour, which reads:
 \bea
 b_1 &=& C_1exp\left[\frac{2(d-Q_0)t}{3}\right] \nonumber \\
 \phi_1 &=& \frac{C_1}{4}exp\left[\frac{2(d-Q_0)t}{3}\right] \nonumber \\
 T_0 &=& C_2exp\left[(A-B)t\right] \nonumber \\
 T_1 &=& C_2(A-B)exp\left[(A-B)t\right], \nonumber \\
 Q_1 &=& 0
 \label{linearsol}
 \eea
 where $A=(4d-Q_0)/6$ and  $B=\sqrt{(4d-Q_0)^2+48}/6$. Note that the
 constant of integration $C_1$ has to be positive. Note also that
 the matter field vanishes exponentially with time, a
 characteristic shared also by the full solution as we shall see later on.
 \pr
 Now  at the first step in the iteration
  the solution is given by:
 \be
 \vec{x}(t) = {\bf Y}(t) \vec{C} + {\bf Y}(t) \int_{t_0}^{t}ds {\bf Y}(t)
 {\bf Y}^{-1}(s) \vec{{\bf F}} \left[{\bf Y}(s) \vec{C} \right],
 \label{firstiter}
 \ee
 where ${\bf Y}(t) \vec{C}$, is the solution of the linear part of 
 the system while ${\bf Y}(t_0) \vec{C}$ is the set of initial conditions.
 Even from the first step of the iteration one can see that there
 are regions of the parameters (e.g $C_1 \gg C_2$) for which the
 dilaton field remains practically constant and the scale factor of
 the metric increases exponentially. We remind the reader that
 this is a common feature of all inflationary scenaria,
 where the r\^{o}le of the inflaton is played in this
 case by the dilaton field. Of course
 the solutions at this order although instructive are not full
 solutions of the system. So we proceed to present full
 numerical solutions. In order to get these solutions we use
 the iteration up to second order. In this way we take expressions
  for the fields which are almost exact asymptotically due to the
  convergence of the iteration. Then we use the values of
  these expressions at a certain point (in time) as initial
  conditions and we let the system evolve numerically.
  \pr
  One of the basic features of all solutions is the existence
  of an initial singularity ($t \rightarrow -\infty$). 
  Of course our consideration starts
  immediately after the singularity since a proper treatment of
  true initial conditions of the Universe is an issue which
  has to take into account the full quantum gravity effects.   
Indeed, we do not expect a proper string theoretic 
cosmological model to be described only in terms of an 
effective field theory based on the low-string-level fields. 
In the context of our non-critical 
string scenario, the existence of an initial singularity, 
describes quite naturally
the effects of higher-string level degrees of freedom,
including the quantum-mechanical (discrete) ones, responsible 
for the non-criticality of the effective field theory~\cite{emn}.
The effects of such modes are expected to be 
strong at very early stages of the Universe. 
Moreover, higher curvature terms, as well as higher order 
string loop corrections
should also be taken into account at such early stages, 
given that the 
string coupling $g_s$ grows infinitely strong near the initial 
singularity.
  \pr
  We now note that 
  the signature of the inflationary era is the sign of the
  difference $Q(t) - \dot{\Phi}(t)$ as is already mentioned.
  This difference becomes negative for a certain period
  provided that the density of the matter field ($T$) becomes
  weak immediately after the singularity.  In particular in the
  inflationary era it is of the order of magnitude of
  the asymptotic value of $Q$, ($Q_0$).
  In the figures that follow we present a particular solution
  in which the abovementioned characteristics become more clear.
  \pr
  In figure 1 the scale factor ($exp[b(t)]$) is plotted. It is
  clear that it exhibits an exponential growth period. The exit
  from the inlfation comes by a power-law scale factor which
  settles down to a flat space-time metric. From figure 2 it is
  clear that the difference  $Q(t) - \dot{\Phi}(t)$ being
  negative (certifying the inflation) at early times turns to
  positive values indicating thus the exit from inflation.
  In figure 3 the dilaton field is plotted. Note that it
  remains practically constant during inflation, whilst 
  at later times it {\it decreases linearly} with the cosmic time.
  This latter feature is very important, as it implies that 
the string coupling $g_s \sim e^{\Phi} \rightarrow 0$, as $t \rightarrow
\infty$, and therefore the tree-level effective action (\ref{actioneff})
is {\it sufficient} to describe the transition from the 
inflationary to post-inflationary era. 
This feature is {\it due to the 
supercriticality } of the initial stringy configurations, $C > 25$. 
It is the opposite situation
from what is happening in critical strings~\cite{veneziano,ven2},
thereby making our super-critical (Liouville) string model a viable 
model to describe  `graceful exit' from the inflationary era.  
As mentioned 
previously, of course,  
our tree-level effective-field theory description 
breaks down for very early stages,
near the initial singularity. 
  
\pr
In figure 4 we show the matter field ($T$). We see here that
  at the beginning of the inflation the value of the matter
  field is of the order of the asymptotic value of $Q$. Solutions
  with matter density one order of magnitude bigger than
  $Q_0$ do not have inflationary era. This can be understood
  since in a condensed enough Universe gravitational forces
  prevent exponential expansion.
 In figure 5 we show the evolution of $Q$. We see that 
 during the inflationary period it is almost linerly dependent
 on time and it relaxes to a constant (non-zero value)
 asymptotically. We also note that $Q(t)$ {\it changes sign} 
 at the end of inflation. 
Finally in figure 6 we show, for completeness, 
the time dependence of the 
`running' central charge deficit $Q^2(t)$. 
Note that, in contrast to 
the inflationary phase where $Q^2(t)$ decreases rapidly with time, 
the 
post-inflationary
period is characterized by an increasing with time $Q^2(t)$
before the latter settles to its final (equilibrium) constant value
asymptotically.
This is in agreement with the fact that the Liouville mode is 
a non-unitary world-sheet field
for supercritical strings~\cite{aben}. Thus, the  
conditions of ref. \cite{zam} for a monotonic
decrease of the running central charge are not valid. 
However, as we see from figure 6, there is an  overall 
decrease of the central charge deficit 
during the flow 
from its initial 
(`near the singularity') to the final (equilibrium) value. 
We point out that a similar situation, 
where the central charge
decreases overall, but oscillates before settling to its
non-trivial world-sheet fixed point value, also characterizes
the dilaton cosmology of \cite{schmid}.
\pr
The overall decrease of $Q^2(t)$ in Liouville strings is 
expected 
on general grounds~\cite{kutasov}, given the 
connection of the irreversibility of the world-sheet renormalization
group flow with `loss of information' associated 
with stringy modes having world-sheet momenta 
beyond the ultraviolet cut-off of the effective theory~\cite{zam,emn}.
This world-sheet cut-off should not be confused with a
space-time cut-off. However, one may find a proper mapping 
to a target-space ultraviolet scale,  
as a 
result of the embedding of the world sheet in a target space time. 
It is in this sense that the non-criticality of the effective 
string describes information loss due to 
stringy modes unobservable in the context of the low-energy effective 
field theory
describing cosmological observations.

\subsection{Solutions attracted by a constant dilaton fixed point}

We next turn our attention to demonstrating the existence 
of solutions to (\ref{system}) 
which are characterized by an asymptotic
vanishing of the central 
charge deficit $Q^2 (t\rightarrow \infty) \rightarrow 0$. 
Such solutions are known to characterize certain two-dimensional 
black hole 
models~\cite{gang6}, and it may be the case that exact conformal 
field theories exist which also describe 
cosmological models. From the inflationary scenaria view point
such solutions are important in that they are characterized by 
a relaxing to zero cosmological constant in target space~\cite{gang6},
something which might be  a phenomenological requirement
when one extends such scenaria to four dimensional space times. 
\pr
The method adopted previously is not convenient to find these
solutions. The main reason is that if we set $d=Q_0=0$ in 
(\ref{separate})
the linear part of the system (\ref{system}) becomes trivial and does 
not permit the iteration (\ref{itersol}). 
Nevertheless we can see that 
there exist such solutions with the desired features (inflationary
era and asymptotic flatness), using a different method which we
describe shortly in the following.
\pr
To this end we
first note that, upon using the first of 
equations (\ref{system}),  
the relation 
\be
Q=-\dot{b}
\label{relqb}
\ee
solves
the last of these equations.  
Hence, in the following
we shall seek solutions satisfying the relation (\ref{relqb})
between the central charge and the field $b$. 
From the solutions we
have already obtained in our analysis above 
we see that this relation holds immediately after
the (initial) singularity. 
If we assume this feature to 
characterize also the asymptotic region ($t \rightarrow
\infty)$, then we observe that it is possible 
to enforce the central charge deficit to {\it vanish asymptotically},
$Q^2 (t) \rightarrow 0$. The
second fact that we infer from the solutions we have obtained 
above
is that
the field $b$ is monotonic in time. If we assume this to hold in the 
case of asymptotically vanishing central charge deficit as well, then 
we can consider the fields $\Phi$ and $T$ being functions
of $b$. 
Anticipating flat space 
time solutions asymptotically ($t \rightarrow \infty$),
we shall concentrate on the case of weak $b \rightarrow 0$ field, 
for which a perturbative
expansion of the solutions in powers of $b$ is valid.
Indeed, 
rewriting the system of equations (\ref{system}) 
in terms of the new variable, $b$, 
we can find a series solution for these fields. 
We omit 
the details for brevity, and we only state the final result for 
the solution:

\bea
T(b) = c_2\left\{-b - \frac{b^2}{6} - \frac{55+12c_2^2}{216}b^3 - 
\frac{109+60c_2^2}{2592}b^4 \right\}+ O(b^5) \nonumber \\
\Phi(b) = \frac{b}{4} + \frac{1+4c_2^2}{16}b^2 + \frac{c_2^2}{18}b^3
 + \frac{212c_2^2-3}{3456}b^4 + O(b^5).
\label{seriessol}
\eea
This solution gives asymtotically vanishing matter density 
and asymptotically
constant dilaton. This constant can be set equal to zero since our
equations are insensitive in constant dilaton shifts. 
For the field $b$ itself we find the equation:
\be
\dot{b}^2(t) = b^2(t) + a_3(c_2)b^3(t) + a_4(c_2)b^4(t)
\label{seriesb}
\ee
where the constants $a_3(c_2)$, $a_4(c_2)$ are complicated expressions
of $c_2$ which can be found from the expression:
\be
\dot{b}^2 = \frac{2T^2(b)}{\frac{13}{12} + \frac{7}{6}\phi'(b)
 - 4\phi'^2(b) + \frac{3}{2}T'^2(b)}.
\ee
Solving now the equation (\ref{seriesb}) for b as a function of
time the solution (relevant for our purpose) is of the form:

\be
b(t) = -\frac{4e^{t+c_0}}{a_3^2-4a_4+2a_3e^{t+c_0}+e^{2(t+c_0)}}.
\label{inflatb}
\ee
We therefore see clearly that there are solutions 
which asymptotically
in time ($t \rightarrow 0$) 
lead to flat space with constant dilaton and vanishing
central charge deficit. 
Already from the above approximate expression we 
can distinguish two major cases: (i) if we consider
solutions with $a_4>0$ the expression in (\ref{inflatb}) yields
an exponential factor of the form already presented in figure 1.
(ii) On the other hand in the case of $a_4<0$ 
the expression in
(\ref{inflatb}) indicates a
scale factor leading to flat space in both asymptotic regions
with a deep throat in the middle region. Although this latter
class is interesting it lies beyond the scope of this article.
Of course these  expressions
are approximate, but the two classes
of solutions can be confirmed numerically, upon using initial
conditions for the fields consistent with the above approximate
solutions.

\section{Instead of Conclusions}

{}From the analysis 
presented in this article it seems that a simple two-dimensional 
cosmological model, based on super-critical (Liouville) 
strings,
is characterized by phenomenologically acceptable features:  
exponential expansion (inflation), 
`graceful exit' from it, and relaxation towards  an
asymptotically `flat' string universe 
with a 
non-zero
or zero 
constant 
vacuum energy (depending on the initial conditions).  
The most important feature is that 
the 
`graceful exit' from the inflationary 
phase is achieved because of the fact that 
the Liouville $\sigma$-model is 
attracted by the linear- or constant- dilaton 
fixed points (depending on the initial conditions), 
in such a way that the string coupling always remains bounded 
during the inflationary and post-inflationary periods. 
Thus, higher world-sheet topologies do not play a r\^ole 
in the physics of the asymptotic time region $t \rightarrow \infty$. 
This should be contrasted with the 
critical-string situation. 
\pr
The r\^ole of the non-criticality of the string, viewed as a 
non-equilibrium system was crucial. 
In this respect we note that 
in our approach there is non-trivial entropy production~\cite{emn},
determined by the overall decrease of 
the effective (`running') central charge of the theory, $Q^2(t) >0$, 
during the flow 
towards a non-trivial fixed point. The 
entropy change expresses
the amount of information carried by (string) modes 
whose world-sheet `momenta' lie above   
the ultraviolet cut-off scale 
of the effective theory.
The issue of precise estimates of the entropy production in the 
context of non-critical strings is left for future investigations. 
\pr
An additional important feature of our approach was 
the r\^ole of matter in inducing the above-described
temporal evolution of the non-critical string universe.
It is not 
clear to us whether the existence of an initial singularity,
which seems to characterize the solutions,
is an inevitable feature of all such 
non-critical string 
scenaria, even in higher 
dimensions, or just a peculiarity of the two-dimensional 
toy model.
It would be interesting to attempt to extend this analysis
to higher dimensional theories, based on non-critical strings, 
including finite temperature considerations. This 
would allow for a study of more realistic inflationary scenaria,
including the issue of reheating, a currently ``hot''' subject.
\pr
It should be stressed once again that above 
we have worked in the so-called $\sigma$-model
(string) frame, in which lengths have been measured
in string `rods'. 
Thus the inflationary scenaria we have found
represent a true expansion of the stringy universe. 
In two-dimensions this is the only frame,
and one is free from the ambiguities characterizing the 
four dimensional case, where a simple linear dilaton 
solution in the Einstein frame 
may be interpreted~\cite{sanchez} 
as an {\it equilibrium} 
solution corresponding to a non-expanding constant 
universe, provided the measurement of distances is done 
in `string rods'. In such a case one expects 
that 
inflationary, or in general expanding-universe, scenaria 
would be described by more complicated space time configurations. 
\pr
It should also be born in mind that, 
at present, our 
results are preliminary, 
and one cannot make definite claims
regarding the `exit problem' from the 
inflationary phase of (non-critical) string theory. 
Many issues, such as the r\^ole of metric fluctuations
on the `exit' phase, 
entropy estimates and bounds in our non-critical 
string universe, reheating etc, 
are left open. These will hopefully constitute topics of future work. 
However, we believe
that the current results are sufficiently interesting to 
encourage 
further studies
of inflationary scenaria based on non-critical (non-equilibrium) 
Liouville strings. 

\section*{Acknowledgements}
 
We wish to thank V. Mitsou for her assistance in formatting and editing  
the figures. 
The work of N.E.M. is partially supported by a P.P.A.R.C. (U.K.) 
Advanced Research Fellowship.

\newpage

\begin{figure}[htb]
\begin{center}
\epsfxsize=5in
\bigskip
\centerline{\epsffile{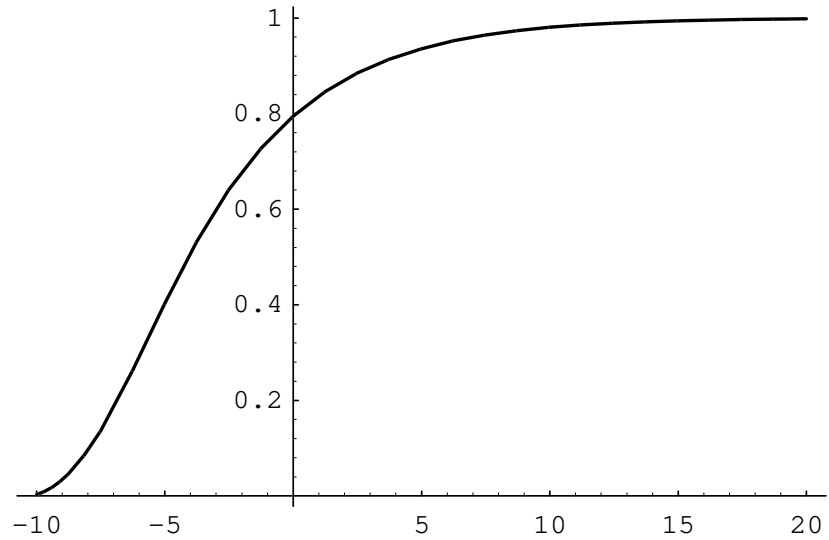}}
\end{center}
\caption{The scale factor $e^{b(t) }$ is plotted versus the cosmic time $t$,
in our supercritical-string-inspired cosmological model.
The system is characterized by a period of exponential growth 
(inflation), which in the units of the figure lies 
in the time interval between -10 and -5, succeeded by a 
period of power-law expansion, and,      
eventually, a stationary phase for $t > 10$.} 
\bigskip
\label{fig1}
\end{figure}

\newpage

\begin{figure}[htb]
\epsfxsize=5in
\bigskip
\centerline{\epsffile{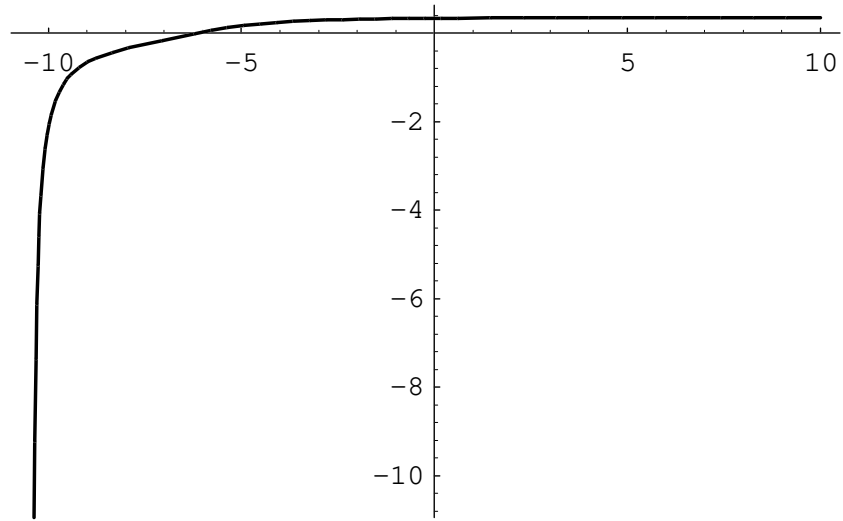}}
\vspace{.8in}
\caption{The difference $Q(t) - {\dot \Phi }$ versus the cosmic time $t$, 
whose sign is  
crucial for the existence of the inflationary period. 
During the inflationary period this quantity is negative. 
Then, its sign changes,
signalling
the exit from the inflationary era.}  
\bigskip
\label{fig2}\end{figure}

\newpage

\begin{figure}[htb]
\epsfxsize=5in
\bigskip
\centerline{\epsffile{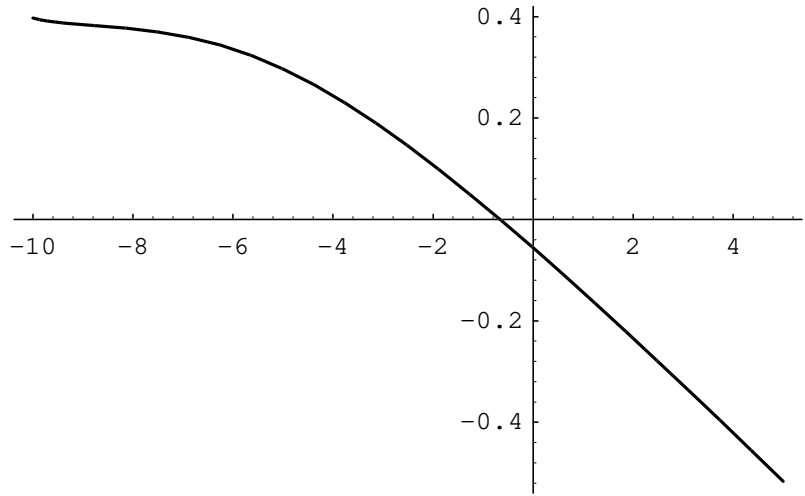}}
\caption{The dilaton configuration versus the cosmic time.
The dilaton starts from positive values. Near the initial 
singularity (not exhibited in the figure) the dilaton 
approaches positive infinity. Its value drops sharply  
towards the inflationary era. During inflation
the dilaton remains finite  and almost constant (positive).
it changes sign during the exit period,  
and 
becomes linearly decreasing with cosmic time asymptotically. } 
\bigskip
\label{fig3}\end{figure}

\newpage

\begin{figure}[htb]
\epsfxsize=5in
\bigskip
\centerline{\epsffile{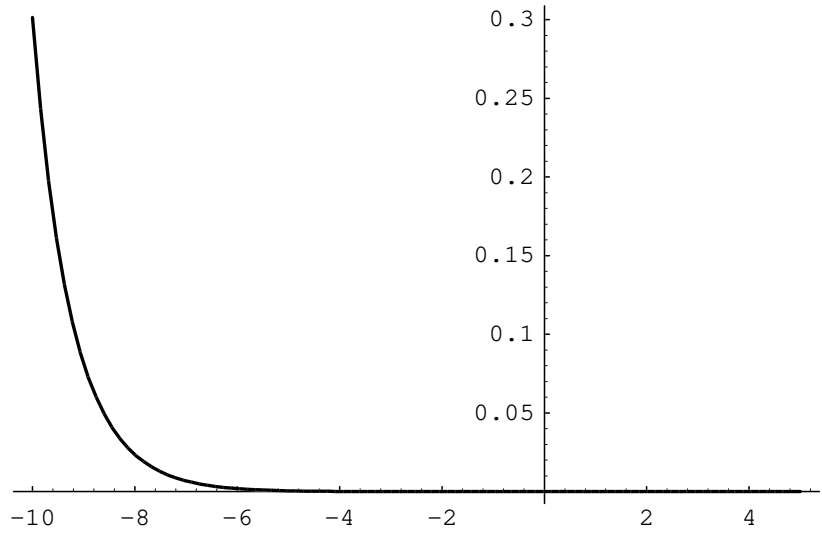}}
\caption{The matter field (Tachyon) as a function of 
the cosmic time.
During the inflationary period it drops sharply from 
large values, of order of the central charge deficit 
$Q$, to practically zero value.} 
\bigskip
\label{fig4}\end{figure}

\newpage

\begin{figure}[htb]
\epsfxsize=5in
\bigskip
\centerline{\epsffile{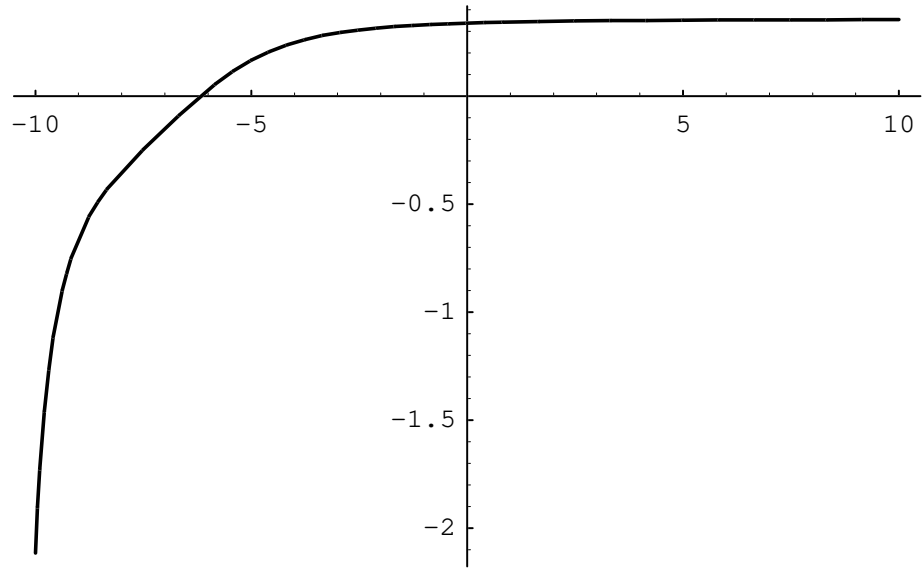}}
\caption{The 
evolution of $Q(t)$ versus the cosmic time $t$. 
Note that the inflationary
period is characterized by an almost linear dependence on time.}
\bigskip
\label{fig5}
\end{figure}

\newpage

\begin{figure}[htb]
\epsfxsize=5in
\bigskip
\centerline{\epsffile{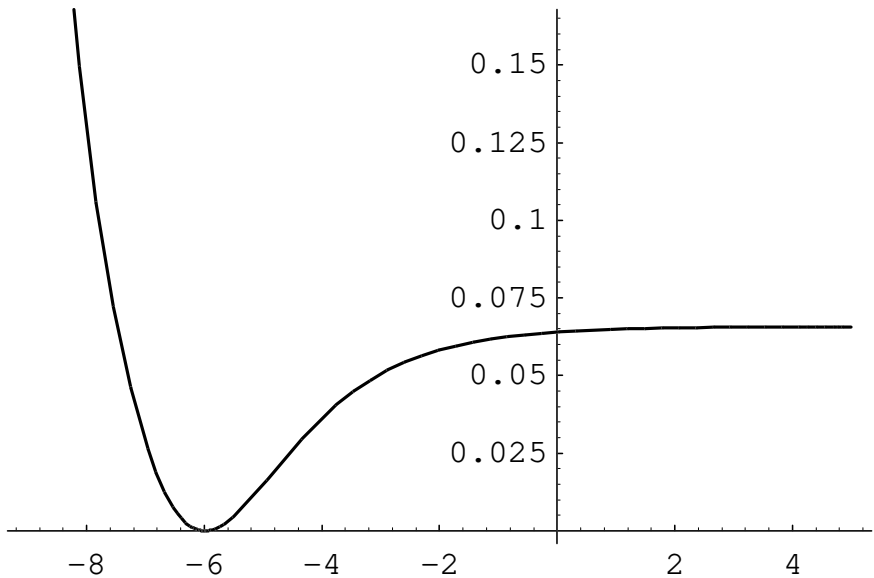}}
\caption{The 
evolution (versus the cosmic time $t$) 
of the central charge deficit 
$Q^2(t) \ge 0$ in our 
supercritical string model. 
Note that $Q^2(t)$ 
decreases rapidly with $t$ during inflation.
In contrast, 
the post-inflationary
period is characterized by an increasing with time central charge
deficit until the latter settles to its final (equilibrium) value.
The increase of $Q^2(t)$ 
is in agreement with the fact that the Liouville mode is 
a non-unitary world-sheet field. However, there is an  overall 
decrease of the central charge deficit during the flow from its initial 
(`near the singularity') to a final (equilibrium) value, as expected 
on general grounds.}
\bigskip
\label{fig6}
\end{figure}

 \end{document}